\documentclass[a4paper,12pt]{revtex4}
\usepackage{amsfonts,amsmath,amssymb}
\usepackage{amssymb}
\usepackage{epstopdf}

\usepackage{graphicx}

\begin{document}

\font\msytw=msbm10 scaled\magstep1

\let\a=\alpha \let\b=\beta  \let\g=\gamma  \let\d=\delta \let\e=\varepsilon
\let\z=\zeta  \let\h=\eta   \let\th=\theta \let\k=\kappa \let\l=\lambda
\let\m=\mu    \let\n=\nu    \let\x=\xi     \let\p=\pi    \let\r=\rho
\let\s=\sigma \let\t=\tau   \let\f=\varphi \let\ph=\varphi\let\c=\chi
\let\ps=\psi  \let\y=\upsilon \let\o=\omega\let\si=\varsigma
\let\G=\Gamma \let\D=\Delta  \let\Th=\Theta\let\L=\Lambda \let\X=\Xi
\let\P=\Pi    \let\Si=\Sigma \let\F=\Phi    \let\Ps=\Psi
\let\O=\Omega \let\Y=\Upsilon

\def\\{\hfill\break} \let\==\equiv
\let\txt=\textstyle\let\dis=\displaystyle

\let\io=\infty \def\Dpr{\V\dpr\,}
\def\aps{{\it a posteriori\ }}\def\ap{{\it a priori\ }}
\let\0=\noindent\def\pagina{{\vfill\eject}}
\def\bra#1{{\langle#1|}}\def\ket#1{{|#1\rangle}}
\def\media#1{{\langle#1\rangle}}
\def\ie{{i.e. }}\def\eg{{e.g. }}
\let\dpr=\partial \def\der{{\rm d}} \let\circa=\cong
\def\arccot{{\rm arccot}}

\def\ux{{\underline x}}
\def\uy{{\underline y}}
\def\uz\def\ux{{\underline z}}

\def\sech{{\rm sech}}
\def\PPP{{\cal P}}\def\EE{{\cal E}}\def\MM{{\cal M}}
\def\CC{{\cal C}}\def\FF{{\cal F}} \def\HHH{{\cal H}}\def\WW{{\cal W}}
\def\TT{{\cal T}}\def\NN{{\cal N}} \def\BBB{{\cal B}}\def\III{{\cal I}}
\def\RR{{\cal R}}\def\LL{{\cal L}} \def\JJ{{\cal J}} \def\OO{{\cal O}}
\def\DD{{\cal D}}\def\AAA{{\cal A}}\def\GG{{\cal G}} \def\SS{{\cal S}}
\def\KK{{\cal K}}\def\UU{{\cal U}} \def\QQ{{\cal Q}} \def\XXX{{\cal X}}

\def\hh{{\bf h}} \def\HH{{\bf H}} \def\AA{{\bf A}} \def\qq{{\bf q}}
\def\VV{{\cal V}}
\def\BB{{\bf B}} \def\XX{{\bf X}} \def\PP{{\bf P}} \def\pp{{\bf p}}
\def\vv{{\bf v}} \def\xx{{\bf x}} \def\yy{{\bf y}} \def\zz{{\bf z}}
\def\aaa{{\bf a}}\def\bbb{{\bf b}}\def\hhh{{\bf h}}\def\II{{\bf I}}
\def\ii{{\bf i}}\def\jj{{\bf j}}\def\kk{{\bf k}}\def\bS{{\bf S}}
\def\Vm{{\bf m}}\def\Vn{{\bf n}}

\def\ol#1{{\overline #1}}

\def\to{\rightarrow}
\def\la{\left\langle}
\def\ra{\right\rangle}

\def\RRR{\hbox{\msytw R}} \def\rrrr{\hbox{\msytww R}}
\def\rrr{\hbox{\msytwww R}} \def\CCC{\hbox{\msytw C}}
\def\cccc{\hbox{\msytww C}} \def\ccc{\hbox{\msytwww C}}
\def\NNN{\hbox{\msytw N}} \def\nnnn{\hbox{\msytww N}}
\def\nnn{\hbox{\msytwww N}} \def\ZZZ{\hbox{\msytw Z}}
\def\zzzz{\hbox{\msytww Z}} \def\zzz{\hbox{\msytwww Z}}
\def\SSS{\hbox{\msytw S}} \def\bbbb{\hbox{\msytww B}}
\def\ssss{\hbox{\msytww S}} \def\sss{\hbox{\msytwww S}}
\def\TTT{\hbox{\msytw T}} \def\tttt{\hbox{\msytww T}}
\def\ttt{\hbox{\msytwww T}}\def\MMM{\hbox{\euftw M}}
\def\QQQ{\hbox{\msytw Q}} \def\qqqq{\hbox{\msytww Q}}
\def\qqq{\hbox{\msytwww Q}}

\def\vvv{\hbox{\euftw v}}    \def\vvvv{\hbox{\euftww v}}
\def\vvvvv{\hbox{\euftwww v}}\def\www{\hbox{\euftw w}}
\def\wwww{\hbox{\euftww w}}  \def\wwwww{\hbox{\euftwww w}}
\def\vvr{\hbox{\euftw r}}    \def\vvvr{\hbox{\euftww r}}

\def\ul#1{{\underline#1}}
\def\Sqrt#1{{\sqrt{#1}}}
\def\V0{{\bf 0}}
\def\defi{\,{\buildrel def\over=}\,}
\def\lhs{{\it l.h.s.}\ }
\def\rhs{{\it r.h.s.}\ }

\let\wt=\widetilde\def\AAA{{\cal A}}\let\oo=\Bo\let\nn=\Bn
\let\aaa=\Ba\let\pps=\Bps\def\hhh={\V h}\def\bbb{{\V b}}
\let\bb=\Bb\def\ss{\ul{\s}}

\def\RRR{\hbox{\msytw R}} \def\rrrr{\hbox{\msytww R}}
\def\rrr{\hbox{\msytwww R}} \def\CCC{\hbox{\msytw C}}
\def\cccc{\hbox{\msytww C}} \def\ccc{\hbox{\msytwww C}}
\def\NNN{\hbox{\msytw N}} \def\nnnn{\hbox{\msytww N}}
\def\nnn{\hbox{\msytwww N}} \def\ZZZ{\hbox{\msytw Z}}
\def\zzzz{\hbox{\msytww Z}} \def\zzz{\hbox{\msytwww Z}}
\def\TTT{\hbox{\msytw T}} \def\tttt{\hbox{\msytww T}}
\def\ttt{\hbox{\msytwww T}}
\def\QQQ{\hbox{\msytw Q}} \def\qqqq{\hbox{\msytww Q}}
\def\qqq{\hbox{\msytwww Q}}
\def\SSS{\hbox{\msytw S}} \def\BBBB{\hbox{\msytw B}}

\let\ul=\underline\def\hh{{\V h}}
\def\cfr{{cf. }}\let\ig=\int
\def\Tr{{\rm Tr}}
\def\dist{{\rm dist}}

\def\gtopl{\hbox{\msxtw \char63}}
\def\ltopg{\hbox{\msxtw \char55}}

\def\be{\begin{equation}}
\def\ee{\end{equation}}
\def\bea{\begin{eqnarray}}\def\eea{\end{eqnarray}}
\def\bean{\begin{eqnarray*}}\def\eean{\end{eqnarray*}}
\def\bfr{\begin{flushright}}\def\efr{\end{flushright}}
\def\bc{\begin{center}}\def\ec{\end{center}}
\def\ba#1{\begin{array}{#1}} \def\ea{\end{array}}
\def\bd{\begin{description}}\def\ed{\end{description}}
\def\bv{\begin{verbatim}}\def\ev{\end{verbatim}}
\def\nn{\nonumber}
\def\Halmos{\hfill\vrule height10pt width4pt depth2pt \par\hbox to \hsize{}}
\def\pref#1{(\ref{#1})}
\def\Dim{{\bf Dim. -\ \ }} \def\Sol{{\bf Soluzione -\ \ }}
\def\virg{\quad,\quad}
\def\bsl{$\backslash$}
\newtheorem{result}{Result}[section]
\newtheorem{lemma}{Lemma}[section]
\newtheorem{proposition}{Proposition}[section]
\newtheorem{theorem}{Theorem}[section]
\newtheorem{proof}{Proof}[section]
\newtheorem{remark}{Remark}[section]
\newtheorem{corollary}{Corollary}[section]
\renewcommand{\theequation}{\arabic{section}.\arabic{equation}}


\begin{center}
{\Large\bf{Quantum Quench dynamics in Non-local Luttinger Model: Rigorous Results}}
\vskip1cm



Zhituo Wang\\ {\it Institute for Advanced Study in Mathematics, Harbin Institute of Technology,\\ 150006, Harbin, China, \\ Research Center for Operator Algebras, East China Normal University,}\\ Email:wzht@hit.edu.cn
\end{center}

\begin{abstract}
We investigate, in the Luttinger model with fixed box potential, the time evolution of an inhomogeneous state prepared as a localized fermion added to the noninteracting ground state. We proved that, if the state is evolved with the interacting Hamiltonian, the averaged density has two peaks moving in opposite directions, with a constant but renormalized velocity. We also proved that a dynamical `Landau quasi-particle weight' appears in the oscillating part of the averaged density, asymptotically vanishing with large time. 
The results are proved with the Mattis-Lieb diagonalization method. A simpler proof with the exact Bosonization formulas is also provided. 
\end{abstract}


\maketitle

\renewcommand{\thesection}{\arabic{section}}

\section{Introduction}

Recent experiments on cold atoms \cite{B1} have motivated increasing interest
in the dynamical properties of many body quantum systems which are {\it closed} and isolated from any
reservoir or environment \cite{P}.
Nonequilibrium properties can be investigated by {\it quantum quenches}, in which
the system is prepared in an eigenstate of the non-interacting Hamiltonian and its subsequent time evolution driven by an interacting many-body Hamiltonian is observed.
As the resulting dynamical behavior is the cumulative effect of the interactions between an infinite
or very large number of particles, the computation of local observables averaged over time-evolved states poses typically
great analytical difficulties; therefore, apart for some analysis in two dimensions (see, for instance \cite{Rig5, Rig6}), the problem is mainly studied in one dimension \cite{A}-\cite{Ha4}. A major difference with respect to the {\it equilibrium} case relies on the fact that in such a case
a form of {\it universality} holds, ensuring that a number of properties are essentially {\it insensitive}
to the model details. At non-equilibrium the behavior depends instead on model details; for instance 
integrability in spin chains dramatically affects the non equilibrium behavior
\cite{C2}, \cite{A1},\cite{A2} while it does not alter the $T=0$ equilibrium properties \cite{M1}.
This extreme sensitivity to the details or approximations asks for a certain number of analytical exact results
at non-equilibrium, to provide a benchmark for 
experiments or approximate computations.

One of the interacting Fermionic system where non-equilibrium properties can be investigated
is the Luttinger model \cite{L, T} (see also \cite{ML,vieri1, MM}), which provides a great number of information in the equilibrium case. In the Luttinger model model the quadratic dispersion relation of the non relativistic fermions is replaced with a linear dispersion relation, leading to the "anomaly" in the distribution of the ground states density. This anomaly is proved to be universal for a large class of one dimensional Fermionic system, called the Luttinger liquid \cite{Ha}.  
Luttinger model became of great interest in mathematical physics ever since the exact solutions founded by Mattis-Lieb \cite{ML} and is a key to investigate the mathematical properties of condensed matter physics.

It is important to stress that there exist two versions of this model, 
the {\it local Luttinger model} (LLN)
and the {\it non local} Luttinger model (NLLM); in the former a local delta-like interaction is present while in the latter the interaction is short ranged but non local. 
The finite range of the interaction plays as an ultraviolet cut-off. At equilibrium such two models are often confused as they have similar behavior, due to the above mentioned insensitivity to model details; there is however no reason to expect that 
this is true also at non equilibrium. It should be also stressed that the LLM is
plagued by {\it ultraviolet divergences} typical of a  QFT and an {\it ad-hoc} regularization is necessary to get physical predictions; the short time or distance behavior depends on the chosen regularization.

In this paper we study the evolution of {\it inhomogeneous} states in the non-local Luttinger Model with a fixed {\it box potential}, with the Mattis-Lieb diagonalization method, which was proved to be mathematically rigorous (\cite{vieri1,MM}). Then we perform rigorous analysis of the asymptotic behavior in the infinite volume limit. The main result shows that (see Theorem \ref{mainthm}),
when the interaction is turned on, the dynamics is ballistic with a constant but renormalized velocity, and the interaction produces a 
dynamical `Landau
quasi-particle weight' in the oscillating part, asymptotically vanishing with time. The expressions we get do not require any ultraviolet regularization, and correctly capture also the short time dynamics. We also invite the physically oriented reader to read this article along with a short letter \cite{maswang}, in which we studied the quench dynamics of non-local Luttinger model but without giving full details of the proof. In the current article we put full details of the proof and specialize to the box potential, for which the change of velocity due to the many-body interaction is more transparent; we provide also a simpler proof of the main theorem with the exact Bosonization formulas.

The quantum quench of {\it homogeneous}
states in the NLLM was derived in \cite{MG3},\cite{Me}, in which steady states were found. However mathematical rigor is lacking in these work. The quenched evolution of the NLLM prepared in domain wall initial state was studied in \cite{lebow1} and
the universality of the quantum Landauer conductance for the final states was proved, in a mathematically rigorous way.

The plan of the paper is the following.
We introduce the NLLM with box potential in \S II. 
In \S III we prove Theorem \ref{mainthm} with the Mattis-Lieb
diagonalization method. Some details of the proof are presented in the Appendix. The proof of Theorem \ref{mainthm} based
on the Bosonization method is given in \S IV. 

\section{The Luttinger model and main results}
\subsection{The Luttinger model with box potential}
The non-local Luttinger model (NLLM) is defined by the
Hamiltonian: 
\bea H_\l&=&\int_{-L/2}^{L/2}dx\ i\
v_F(:\psi^+_{x,1}\partial_x \psi^-_{x,1}:- :\psi^+_{x,2}\partial_x
\psi^-_{x,2}:)\nn\\
&&\quad +\l \int_{-{L\over 2}}^{L\over 2} dx dy\ v(x-y)
:\psi^+_{x,1} \psi^-_{x,1}: :\psi^+_{y,2}\psi^-_{y,2}: 
\eea
where $\psi^\pm_{x,\o}=\frac{1}{ \sqrt{L}}\sum_k a_{k,\o} e^{\pm i
k x}$, $\o=1,2$, $k=\frac{2\pi n}{ L}$, $n\in N$ are fermionic
creation or annihilation operators, $::$ denotes Wick ordering and
$v_F$ is the Fermi velocity. We are choosing units so that $v_F=1$.
The two-body interaction potential $v(x-y)$ is given by:
\be v(x-y)=\frac{\sin (x-y)}{x-y},\ee
whose fourier transform reads:
\be
v(p)=\bigg\{\begin{matrix} v_0\quad for\ p\le 1,\\0\quad for\ p>1.\end{matrix}\label{box}
\ee
The potential $v(x)$ or $v(p)$ is also called the box potential and $v_0$ is called the strength of $v(p)$. Equilibrium Luttinger model with box potential was first considered in
\cite{theumann}. 

In the Fourier space the Luttinger Hamiltonian can be written as
\bea
H&=&H_0+V=\sum_{k>0} k[(a^+_{k,1} a^-_{k,1}+a^-_{-k,1}a^+_{-k,1})+( a^+_{-k,2} a^-_{-k,2}+a^-_{k,2}a^+_{k,2})\nn\\
&+&{\l\over
L}\sum_{p>0}v(p)[\r_1(p)\r_{2}(-p)+\r_{1}(-p)\r_2(p)]+{\l\over L}\
v(0)\ N_1 N_2 \label{halp}\eea
where, for $p>0$,
\bea
&&\r_\o(p)=\sum_k a^+_{k+p,\o}a^-_{k,\o},\quad N_\o=\sum_{k>0}(a^+_{k,\o}a^-_{k,\o}-a^-_{-k,\o}a^+_{-k,\o}).
\eea
It is well known that Fock space canonical commutation relations don't have a unique representation in a system with 
infinite degree of freedom. So one has to introduce a cutoff function $\chi_\L(k)$ with $\L$ a large positive number such that
$\chi_\L(k)=1$ for $|k|\le\L$ and equals $0$ otherwise and the regularized operators
$\r_\o(p)$ must be thought as $\lim_{\L\to\io }\sum_k
\chi_\L(k)\chi_\L(k+p) a^+_{k+p,\o}a^-_{k,\o}$.

The Hamiltonian $H$ as well as $\r_\o(p)$ can be regarded as
operators acting on the Hilbert space $\HHH$ constructed as
follows. Let $\HHH_0$ be the linear span of vectors obtained by
applying finitely many times creation or annihilation operators on
\be |0>=\prod_{k\le 0}a^+_{k,1}a^+_{-k,2}|vac>\label{gs}. \ee
In this way we get an abstract linear space to which we introduced scalar products between
any pair of vectors. $\HHH$ is defined as the completion of $\HHH_0$ in the scalar product
just introduced. Moreover the operators $H$ and $\r_\o(p)$, regarded as operators on $\HHH$ with domain
$\HHH_0$, are self adjoint.

The basic property of the Luttinger model is the validity of the following anomalous
commutation relations, first proved in \cite{ML}, for $p, p'>0$
\be
[\r_1(-p),\r_1(p')]=[\r_2(p),\r_2(-p')]={pL\over 2\pi}\d_{p,p'}.
\ee

Remark that this commutator acting on the Fock space is not precise due to the infinitely 
many degrees of freedom of the system. So one should introduce a cutoff $\Lambda$ so that 
the commutator:
\be
-\sum_{k=\L+p}^\L a^+_{k,\o}a^-_{k,\o}+\sum_{k=-\L}^{\L-p} a^+_{k,\o}a^-_{k,\o}=
\sum_{k=-\L}^{-\L+p} a^+_{k,\o}a^-_{k,\o}-
\sum_{k=\L-p}^{\L} a^+_{k,\o}a^-_{k,\o}\ .
\ee
on any state of $\HHH$ is equal, in the limit $\L\to\io$, to
${pL\over 2\pi}$.

Moreover one can verify that
\be\label{commu2} \r_2(p)|0>=0\ ,  \quad \r_1(-p)|0>=0\ . \ee
Other important commutation relations (see \cite{ML, langman} for proofs) are as follows:
\bea\label{h0} [H_0,\r_\o(\pm p)]=\pm\e_\o p \r_p(\pm p),\quad\quad [\r_\o,\psi^\pm_{\o,x}]=e^{ip x}\psi^\pm_{\o,x}\ 
\eea
where $\o=1,2$; $\e_\o=1$ for $\o=1$ and $\e_\o=-1$ for $\o=2$.
\subsection{The Mattis-Lieb diagonalization}
The Hamiltonian \eqref{halp} can be diagonalized with the method of Lieb-Mattis 
\cite{ML}, as follows.
First of all we introduce an operator 
\be T=
{1\over L}\sum_{p>0}[\r_1(p)\r_{1}(-p)+\r_{2}(-p)\r_2(p)]\ee 
and write $H=(H_0-T)+(V+T)=H_1+H_2$. Note that $H_1$ is already diagonalized
in that it commutes with $\r_\o$.
The key for the diagonalization of $H_2$ is the introduction of
a bounded operator $S$ acting on the Hilbert space $\cal{H}$:
\be\label{S} S={2\pi\over L}
\sum_{p\not=0}\phi(p)p^{-1}\r_1(p)\r_2(-p),\quad{\rm
tanh}\phi(p)=-{\l v(p)\over 2\pi}. \ee 

Using the following Bogolyubov transformations for the operators $\rho_\o(\pm p)$:
\bea\label{rho} && e^{iS}\r_{1,2}(\pm p)e^{-i
S}=\r_{1,2}(\pm p)\cosh\phi(p)+\r_{2,1}(\pm p)\sinh\phi, \eea
we can easily prove that $H_2$ can be written in diagonal form:
\bea\label{rho1}&& e^{iS} H_2 e^{-i S}=\tilde
H_2\nn\\
&&\quad:={2\pi\over L}\sum_p \sech
2\phi(p)[\r_1(p)\r_1(-p)+\r_2(-p)\r_2(p)]+E_0. \eea

By Formula \eqref{S} we can easily find that the operator $S$ hence the transformation
in \eqref{rho1} is well defined only for $|\lambda v(p)|<2\pi$; The model is instable for
$|\lambda v(p)|>2\pi$.

Define \be D=\tilde H_2-T={2\pi\over L}\sum_p  \s(p)
[\r_1(p)\r_1(-p)+\r_2(-p)\r_2(p)]+E_0, \ee 
we have $ [H_0,D]=0$. The diagonalization formula for the Hamiltonian reads:
 \be\label{diag} e^{iS}e^{iHt} e^{-iS}=e^{i(H_0+D)t}.\ee

\subsection{The time evolution of the one particle state and the main theorem}
Define
\bea\label{time}\psi^\pm_{x,\d}&=&e^{i H_0 t}\psi^\pm_{\o,x} e^{-i H_0 t}\nn\\
&=& {1\over\sqrt{L}}\sum_k a^\pm_{\o,k} e^{\pm i (kx -\e_\o kt)-\d
|k|}, \eea where $\d\rightarrow 0^+$, $\e_1=+,\e_2=-$. 
By direct calculation we find that:
\be\label{free1} <0|\psi^{\e_\o}_{\o, x, \d}\psi^{-\e_\o}_{\o, y,
\d}|0>= {(2\pi)^{-1}\over i\e_\o(x-y)-i(t-s)+ \d}\ . \ee

The relation between the creation or annihilation Fermionic operators and the quasi-particle operators is
\be \psi_{x}=e^{i p_Fx}\psi_{x,1}+e^{-i
p_Fx}\psi_{x,2}\ ,\label{10}
\ee
where $p_F$ is the {\it Fermi
momentum} and we call $e^{i p_Fx}\psi_{x,1}= \tilde\psi_{x,1}$ and $e^{-i
p_Fx}\psi_{x,-1}=\tilde\psi_{x,2}$. In momentum space this simply means that the momentum
$k$ is measured from the Fermi points, that is $c_{k,\o}=\tilde
c_{k+\e_\o p_F,\o}$. The ground state of $H$ is $|GS>=e^{iS}|0>$, where $|0>$ is the ground
state of $H_0$ and the inhomogeneous one particle initial state is given by:
\be\label{initial} |I_{t}>=e^{i H_\l t} (\tilde \psi^+_{1,x}+\tilde
\psi^+_{2,x})|0>. \ee
%
%
Let $n(z)$ be the density operator, which is defined as the limit $\d\to 0, \e\to 0$ of the
following expression:
\bea\label{a3} &&{1\over 2}\sum_{\r=\pm}(\tilde \psi_{1, z+\r\e}^+
\tilde\psi_{2, z, }^- +\tilde\psi_{2, z+\r\e}^+ \psi_{1,
z}^-+ \tilde \psi_{2, z+\r\e}^+ \tilde\psi_{2, z}^-\\
&&\quad\quad +\tilde\psi_{1, z+\r\e}^+ \tilde\psi_{1, z}^- + \tilde\psi_{1,
z+\r\e}^+ \tilde\psi_{1, z}^-  + \tilde\psi_{2, z+\r\e}^+
\tilde\psi_{2, z}^- ).\nn\eea
Note that summing over $\r=\pm$ is the point spitting regularization, which plays the same role as the 
Wick ordering for avoiding divergences. We are interested in the average value of the density operator w. r. t. the 1-particle
initial state \eqref{initial}, formally defined by:
%
\bea\label{wickord}&&G(x,z,t,\d):=<I_t|n(z)|I_t>\\
&&\quad\quad:=\sum_{\o,\o'=1,2}\big[\ \langle0|\tilde\psi_{\o,
x}^- e^{i H t} \tilde \psi_{\o, z+\r\e}^+ \tilde\psi_{\o', z}^- e^{-i H t} \tilde\psi_{\o', x}^+\ |0 \rangle\nn\\
&& \quad\quad\quad+\langle0|\  \tilde\psi_{\o, x}^-  e^{i H t} \tilde \psi_{\o', z+\r\e}^+ \tilde\psi_{\o', z}^- e^{-i Ht} \tilde\psi_{\o, x}^+\  |0 \rangle\ \big]\label{nonrea2}\nn
\eea
\vskip.5cm

As a first step we consider the non-interacting case. Let $|I_{0,t}>:=e^{i H_0 t} (\tilde \psi^+_{1,x}+\tilde
\psi^+_{2,x})|0>$, we have:
\begin{theorem}\label{nonint1}
When $\l=0$, $H=H_0$, we have 
\bea \label{finalbal2} 
&&\lim_{L\to\io}<I_{0,t}|n(z)|I_{0,t}>\\
&=&
\frac{1}{2\pi^2}{\cos 2p_F(x-y)
\over (x-z)^2-t^2}+\frac{1}{4\pi^2}\ [{1\over(
(x-z)-t)^2}+{1\over( (x-z)+t)^2}].\nn
\eea
\end{theorem}

\begin{proof}
We consider first the term with $\o=1$, $\o'=2$. Using the explicit expressions of the
Fermionic operators and taking the limit $\e\to 0$, we can easily find that this term is equal to $e^{2ip_F(x-y)}(4\pi^2)^{-1}[(x-z)^2-t^2]^{-1}
$; a similar result is found for the second term. The third and fourth terms are vanishing as $\sum_{\r}{1\over \r\e}=0$; similarly the last two term
give $(4\pi^2)^{-1}[(x-z)\pm t]^{-2}$. Combine all these terms we can derive Formula \eqref{finalbal2}, hence proved this theorem.
\end{proof}

\begin{remark}
The physical meaning of Theorem \ref{nonint1} is quite clear: when the interaction is turned off, the average of the density is sum of two terms, an oscillating and a non oscillating part (when the particle is added to the vacuum there are no oscillations $p_F=0$). At $t=0$ the density is peaked at $z=x$, where the average is singular. With the time increasing the particle peaks move in the left and right directions with
constant velocity $v_F=1$ (ballistic motion); that is, the average of the density is singular at $z=x\pm  t$
and a "light cone dynamics" is found.
\end{remark}
\vskip.5cm
When we turn on the interaction and let the system driven by the full interacting Hamiltonian,
the ground states and the dynamics will be significantly changed. 
The explicit expression of \eqref{wickord} can be derived with the Mattis-Lieb diagonalization method followed by a rigorous analysis of the asymptotic behavior for $L\rightarrow\infty$ and large $t$. We have
\begin{theorem}\label{mainthm} 
Let the interacting box potential (see \eqref{box}) be turned on in the Hamiltonian, let $\gamma_0=\frac{v_0}{2}$ and $\omega_0=\sqrt{1-\big(\frac{v_0}{2\pi}\big)^2}$. The average of the density operator with respect to the one particle initial state $|I_{\l,t}>$ in 
the limit $L\rightarrow\infty$ reads:
\bea \label{finalbalb} &&\lim_{L\to\io}
<I_{\l,t}|n(z)|I_{\l,t}> \nn\\
&=& \frac{1}{4\pi^2}\ [{1\over( (x-z)-t)^2}+{1\over( (x-z)+t)^2}]+
\frac{1}{2\pi^2}\ {\cos2p_F(x-z)\ e^{Z(t)}\over
(x-z)^2-(\omega_0t)^2} .\eea
where
\bea \label{landau}
Z(t)=\gamma_0\int_0^{1} \frac{dp}{p}(\cos 2\omega_0pt - 1)
\eea
is the Landau quasi particle factor, such that $Z(0)=1$ and
 \be
\exp Z(t)\sim cst (\frac{1}{2\omega_0t})^{\gamma_0}, \ee for $t\ge1$.

\end{theorem}


\section{Proof of Theorem \ref{mainthm}}
We consider first the term: 
\be \langle0|\
\psi_{1, x}^- e^{i H t} \psi_{1, z}^+ \psi_{2, z}^- e^{-i H t}
\psi_{2, x}^+\ |0 \rangle, \label{int1}\ee
and forget the phase factor $e^{\pm ip_F x}$ for the moment for simplicity;
these factors are very easy to restore. The rest of this subsection is devoted to the calculation of \eqref{int1}.

Let $I$ be an identity operator in $\cal{H}$. Using the fact that
$e^{-i\e S}e^{i\e S}=I$ and 
$e^{-iHt}e^{iHs}\vert_{t=s}=I$, we can write \eqref{int1} as 
\bea&&\langle0|\  \psi_{1, x}^-e^{-i\e S}(e^{i\e S}  e^{i H t}e^{-i\e S})(e^{i\e S}  \psi_{1, z}^+ e^{-i\e S})\cdot\label{id1}\\
&&\quad\quad\quad\quad\cdot(e^{i\e S}\psi_{2, z}^- e^{-i\e S})(e^{i\e S}e^{-i H  s}e^{-i\e S})e^{i\e S} \psi_{2, x}^+\  |0 \rangle|_{\e=1,s=t}\nn\eea

\begin{lemma}\label{fact1}
Let $\hat I_1$ be an operator valued function of $\rho_1(\pm p)$ and $\psi^\pm_1$ and  
$\hat I_2$ be an operator valued function of $\rho_2(\pm p)$ and $\psi^\pm_2$,
then we have the following factorization Formula for \eqref{int1}:
\be G_1=I_1 I_2,
\ee
where $I_1=\langle0\vert\hat I_1\vert0\rangle$ and $I_2=\langle0\vert\hat I_2\vert0\rangle$.
\end{lemma}
\begin{proof}
We shall prove this lemma by deriving the explicit expressions of $\hat I_1$ and $\hat I_2$. Using the diagonalization formula
\eqref{diag}, formula \eqref{id1} can be written as:
 \bea\label{main2}
&&\langle0|\  \psi_{1, x}^- e^{-i\e S}  e^{i (H_0+D)  t}  e^{i\e S} \psi_{1, z}^+  e^{-i\e S} e^{-i (H_0+D) t}
e^{-i\e S}\cdot\\
&&\quad\quad\quad \cdot e^{i\e S}e^{i (H_0+D)  s} e^{iS} \psi_{2, z}^-  e^{-iS}  e^{-i (H_0+D)  s} e^{i\e S}\psi_{2, x}^+\  |0 \rangle|_{\e=1,\  s=t}. \nonumber
\eea

Now we consider the term of $e^{i\e S} \psi_{1, z}^+
e^{-i\e S}$. It is a well known result \cite{ML} that: 
\bea e^{i\e
S}\psi^{\mp}_{1, z}e^{-i\e S}=\psi^{\mp}_{1, z}W^{\pm}_{1,z} R^{\pm}_{1,z},\label{ml3}
\eea where
\bea
W_{1,z}^{\pm}&=&\exp\{\mp{2\pi\over L}\sum_{p>0}{1\over p}[\r_1(p)e^{-ipz}-\r_1(-p)e^{ipz}](\cosh\e\phi-1)\}\nn\\
R_{1,z}^{\pm}&=&\exp\{\pm{2\pi\over L}\sum_{p>0}{1\over p}[\r_2(p)e^{-ipz}-\r_2(-p)e^{ipz}]\sinh\e\phi\}.
\eea
Similarly one has
\be e^{i\e
S}\psi^{\mp}_{2, z}e^{-i\e S}=\psi^{\mp}_{2, z}W^{\pm}_{2,z} R^{\pm}_{2,z} \label{ml4}
\ee
where
\bea
W_{2,z}^{\pm}&=&\exp\{\mp{2\pi\over L}\sum_{p>0}{1\over p}[\r_1(p)e^{-ipz}-\r_1(-p)e^{ipz}]\sinh\e\phi\}\nn\\
R_{2,z}^{\pm}&=&\exp\{\pm{2\pi\over L}\sum_{p>0}{1\over p}[\r_2(p)e^{-ipz}-\r_2(-p)e^{ipz}](\cosh\e\phi-1)\}.
\eea

Then we consider the term 
\be
 e^{-i\e S}e^{i (H_0+D) t}W_{1,z}^{-}R_{1,z}^{-} e^{-i (H_0+D)
t} e^{i\e S},
\ee
which, after inserting the identity operator $I=e^{i\e S}e^{-i\e S}$ and $I=e^{-i (H_0+D)
t}e^{i (H_0+D)
t}$, is equal to
\bea 
 [e^{-i\e S}e^{i (H_0+D) t}W_{1,z}^{-}e^{-i (H_0+D) t} e^{i\e S}]\cdot [e^{-i\e S}e^{i (H_0+D) t}R_{1,z}^{-}
e^{-i (H_0+D) t} e^{i\e S}].\label{step3}\eea

Let $f(p,t)$ be an arbitrary regular function, define $\s(p)=\sech2\phi-1$ and $\o(p)=\s(p)+1=\sech2\phi$, we have the following commutation
relation \be\label{sig} [H_0+D,\ \r_\o(\pm p)]=\pm\e_\o p(\s(p)+1)\r_\o(\pm
p),\ \o=1, 2,\ \e_1=+, \e_2=-\ , \ee 
which implies that
\be\label{rho4}
e^{i(H_0+D) t}e^{f(p,t) \r_\o(\pm p)} e^{-i(H_0+D)
t}=e^{e^{\pm\e_\o i(\s+1)pt} f(p,t)\r_\o(\pm p)}. \ee 
Combining the above
formula with \eqref{rho} and \eqref{time} we find that \eqref{step3} can be written as a product of
\bea
&&e^{-i\e S}e^{i (H_0+D) t}W_{1,z}^{\pm}e^{-i (H_0+D) t} e^{i\e S}\nn\\
&=&\exp \pm\frac{2\pi}{L}\sum_p\frac{(\cosh\phi-1)}{p}[\ (\rho_1(-p)\cosh\e\phi-\rho_2(-p)\sinh\e\phi)e^{ipx-ipt(\s+1)}\nonumber\\
&-& (\rho_1(p)\cosh\e\phi-\rho_2(p)\sinh\e\phi)e^{-ipx+ipt(\s+1)}\ ]:=\bar W_{1,z}^{\pm}.
\eea
and
\bea
&&e^{-i\e S}e^{i (H_0+D) t}R_{1,z}^{-}
e^{-i (H_0+D) t} e^{i\e S}\nonumber\\
&=&\exp\pm \frac{2\pi}{L}\sum_p\frac{\sinh\phi}{p}[\ (\rho_2(-p)\cosh\e\phi-\rho_1(-p)\sinh\e\phi)e^{ipy+ips(\s+1)}\nonumber\\
&-& (\rho_2(p)\cosh\e\phi-\rho_1(p)\sinh\e\phi)e^{-ipy-ips(\s+1)}]:=\bar R_{1,z}^{\pm}.
\eea

Using again \eqref{ml3}, \eqref{ml4} and \eqref{rho4}, we have:
\bea\label{1ptbarWR}
&& e^{-i\e S}  e^{i (H_0+D) t}   e^{iS} \psi_{1, z}^+  e^{-iS}
e^{- i (H_0+D) t}e^{i\e S}\nonumber\\
&&=z_aA_{1+}A_{1-}A_{2+}A_{2-} \psi_{1, z t,\d}^+ \tilde W^{-1}_{1
t}\tilde R^{-1}_{1 t} W^{-1}_{1 t\e} R^{-1}_{1 t\e}\hat W^{-1}_{1
t\e}\hat R^{-1}_{1 t\e}, \eea  
and \bea\label{2WR}
&& e^{-iS}  e^{iS}  e^{i H  s} e^{-iS}  e^{iS} \psi_{2, z}^-  e^{-iS}  e^{iS}e^{-i H  s}  e^{-iS} e^{iS}\nonumber\\
&&=z_b \bar W_{2 s\e}\bar R_{2 s\e}\hat W_{2 s\e}\hat R_{2
s\e}\tilde W_2\tilde R_2\psi_{2, z, s,\d}B_{1-}B_{1+}B_{2-}B_{2+},
\eea 
where $\tilde W^{-1}_{1,2, t,\e}$, $\tilde
R^{-1}_{1,2,t,\e}$ and $\hat W^{-1}_{1,2, t,\e}$, $\hat R^{-1}_{1,2, t,\e}$
are operators depending on $\r_{1,2}(\pm p)$, respectively and $z_a$, $z_b$ are functions of $p$. The explicit expressions of the above factors are given in the Appendix.

Then we can easily find that the terms depending on $\r_1(\pm p)$ and $\psi_1^\pm$ are factorized with respect to the terms
depending on $\r_2(\pm p)$ and $\psi_2^\pm$. Let
\bea\label{i1}
I_1&:=&\langle0|\hat I_1|0\rangle\nn\\
&&:=
\langle0|\psi_{1x} A_{1+}A_{1-} \psi_{1, zt}^+\tilde W^{-1}_{1} \bar W^{-1}_{1t} \hat W^{-1}_{1t}\tilde W_{2t}\bar W_{2t} \hat W_{2t}B_{1+}B_{1-}|0\rangle,
\eea
and
\bea\label{i2}
I_2&:=&\langle0|\hat I_2|0\rangle\nn\\
&&:=\langle0|A_{2+}A_{2-} \tilde R^{-1}_{1} \bar R^{-1}_{1} \hat R^{-1}_{1}\bar R_{2} \hat R_{2} \tilde R_{2} \psi_{2, zt}  B_{2+}B_{2-}\psi_{2x}^\dagger |0\rangle,
\eea
and using the fact that $z_a=z_b^{-1}$ 
we have
\be G_1=I_1 I_2,
\ee
So we proved Lemma \ref{fact1}.
\end{proof}

\subsection{Calculation of $I_1$ and $I_2$}\label{calci}
In this part we derive the explicit expressions for $I_1$ and
$I_2$. It is also useful to introduce the following proposition, which can be easily proved using \eqref{h0}:
\begin{proposition}
Let $f(p,t)$ is an arbitrary regular function. Then we have:
\be\label{rho4}
e^{iH_0 t}e^{f(p,t)\ \r_\o(\pm p)} e^{-iH_0 t}=e^{f(p,t)\ e^{\pm\e_\o i(\s+1)pt}\r_\o(\pm p)},\ \o=1, 2;\ \e_1=+, \e_2=-,
\ee
\end{proposition}
The basic idea to calculate $I_1$ and $I_2$ is to use repeatedly the Hausdorff to move the operators $\r_1(-p)$ and $\r_2(p)$ to the right
most of the expressions in \eqref{i1} and \eqref{i2}, and move $\r_1(p)$,
$\r_2(-p)$ to the left most of the above expressions. By formula \eqref{commu2} and its adjoint form we know that these
operator annihilate $|0\rangle$ and $\langle 0|$, respectively; the  survived terms are those independent of $\rho_{1,2}(\pm p)$. 
Setting $\e=1$, we
have: 
\bea\label{I1}
I_1&=&\exp\{\frac{2\pi}{L}\sum_p\frac{1}{p}\ [  ( e^{-ip(\sigma+1)( t+ s)}-1)(2\cosh^2\phi\sinh^2\phi+\cosh^3\phi\sinh\phi )\nonumber\\
&+&(e^{ip(\sigma+1)( t+ s)}-1)\cosh\phi\sinh^3\phi+e^{-ip\sigma  t} (-\cosh^2\phi-\sinh^2\phi) \nonumber\\
&+& e^{ip(x-z)+ip(\sigma+1) s} (\cosh\phi\sinh\phi+\cosh^2\phi)- e^{ip(x-z)+ip s}\nonumber\\
&+& e^{ip(x-z)-ip(\sigma+1) t} (-\sinh\phi-\sinh^2\phi)\
]\}\langle0|\psi_{1x} \psi_{1, z, t, \d}^+|0\rangle, \eea and
\bea\label{I2} I_2&=&\langle0|\psi_{2, z, t, \d}^+\psi_{2x}
|0\rangle\exp \{\frac{2\pi}{L}\sum_p\frac{1}{p} \
[(e^{-ip(\sigma+1)( t+ s)}-1)\cosh\phi\sinh^3\phi\nn\\
& +&(e^{ip(\sigma+1)( t+ s)}-1)(\cosh^3\phi\sinh\phi+2\cosh^2\phi\sinh^2\phi)\nonumber\\
&+&e^{-ip\sigma t} (\cosh^2\phi+\sinh^2\phi)- e^{ip(x-z)-ip t}\nonumber\\
&+& e^{ip(x-z)+ip(\sigma+1) s} (-\cosh\phi\sinh\phi-\sinh^2\phi)\nonumber\\
&+& e^{ip(x-z)-ip(\sigma+1) t} (\sinh\phi+\cosh^2\phi)\  ]\}.\eea

Combining \eqref{I1} with \eqref{I2} and setting $s=t$, we get:
\bea\label{main1}
&&\langle0|\  \psi_{1, x}^-  e^{i H t}  \psi_{1, z}^+ \psi_{2, z}^- e^{-i H  t} \psi_{2, x}^+\  |0 \rangle\label{term1}\\
&&=\langle0|\psi_{1x} \psi_{1, z, t, \d}^+|0\rangle
\langle0|\psi_{2, z, t, \d}^+\psi_{2x} |0\rangle\nn\\
&&\quad\times\exp\sum _p\frac{1}{p}\ \bigg[\ (e^{ip(x-z)+ ip (\s+1)t}
-e^{ip(x-z)+ ipt})\nn\\
&&\quad\quad\quad+(e^{ip(x-z)- ip (\s+1) t}-e^{ip(x-z)- ipt})\nn\\
&&\quad\quad\quad+2\sinh\phi\cosh\phi(\sinh\phi+\cosh\phi)^2(\cos2p(\sigma+1)t-1)\
\bigg].\nonumber \eea
It is useful to derive the asymptotic behavior for the second line in \eqref{main1} and we have:
\bea
&&\lim_{\delta\rightarrow0}\lim_{L\rightarrow\infty}\langle0|\psi_{1x} \psi_{1, z t, \d}^+|0\rangle
\langle0|\psi_{2, z t\d}^+\psi_{2x} |0\rangle=\frac{1}{4\pi^2}\frac{1}{(x-z)^2-t^2}
\eea

With the same method we can derive the explicit expression for the other terms in \eqref{wickord}. Restoring the phase factor $e^{\pm ip_F(x-z)}$ and combine all the terms of \eqref{wickord}, we obtain the following desired result:
\bea \label{finalbalc}&&
<I_{\l,t}|n(z)|I_{\l,t}> =\frac{1}{4\pi^2}\ [{1\over(
(x-z)-t)^2}+{1\over( (x-z)+t)^2}]\label{finald}\\
&&+\frac{1}{4\pi^2}\ {e^{Z(t)}\over (x-z)^2-t^2}\big[\ e^{2i p_F(x-z)}
e^{Q_a(x,z,t)}+e^{-2i p_F(x-z)} e^{Q_b(x,z,t)}\ \big],
\nn \eea
where \be Z(t)=\sum_p {2\over p}\sinh\phi\cosh\phi(\sinh\phi+\cosh\phi)^2(\cos2p(\sigma+1)t-1),\ee
\bea
Q_a&=&\sum_p \frac{1}{p}[(e^{ip(x-z)+ ip (\s_p+1)t}-e^{ip(x-z)+ ip t})\nn\\
&+&(e^{ip(x-z)- ip (\s_p+1)t}-e^{ip(x-z)- ipt})],\nn\\
Q_b&=&\sum_p \frac{1}{p}[(e^{-ip(x-z)+ ip (\s_p+1)t}-e^{-ip(x-z)+ ip t})\nn\\
&+&(e^{-ip(x-z)- ip (\s_p+1)t}-e^{-ip(x-z)- ipt})]\big\}.\label{qab}\eea

\subsection{The asymptotic behavior for $L\rightarrow\infty$} 
In this section we shall derive the asymptotic behavior
of Formula \eqref{finalbalc} in the limit $L\rightarrow\infty$. Using
definitions of the hyper-geometric functions we find that \bea
\sech\phi(p)=\frac{1}{2}\bigg(\frac{1+\frac{
v(p)}{4\pi}}{\sqrt{1+\frac{v(p)}{2\pi}}}-1\bigg), \ \
\cosh\phi=\frac{1}{2}\bigg(\frac{1+\frac{
v(p)}{4\pi}}{\sqrt{1+\frac{v(p)}{2\pi}}}+1\bigg), \eea 
where $v(p)$ is the box potential with strength $v_0$ (see Formula \eqref{box}),
we have the following
expression for the critical exponent: 
\bea\label{crit} \gamma(p)=
2\sinh\phi(p)\cosh\phi(p)(\sinh\phi(p)+\cosh\phi(p))^2
=\frac{v(p)}{4\pi}. \eea 

Taking the limit $L\rightarrow\infty$ means that we should consider the discrete sum over $p$ as integral over continuous variables. We have: 
\bea
Z(t)&=&\int_0^{\infty} \frac{\gamma(p)dp}{p}(\cos 2\omega_0 pt - 1)\nn\\
&=&\gamma_0\int_0^{1} \frac{dp}{p}(\cos 2\omega_0 pt - 1)\nn\\
&=&\gamma_0\int_0^{2\omega_0 t} \frac{d(2\omega_0 p)}{2\omega_0 p}(\cos 2\omega_0 pt - 1),
\eea
where $\gamma_0:=\frac{v_0}{4\pi}$ and
$\omega_0:=\sqrt{1-\big(\frac{v_0}{2\pi}\big)^2}$.
The second line is true is due to the fact that $\gamma(p)=0$ for $p\in(1,\infty]$. Let $y=2\omega_0 pt$ and $w=2\omega_0 t$, $Z(t)$ can be written as:
\bea\label{assmp1}
\gamma_0\int_0^{1} \frac{dp}{p}(\cos 2\omega_0 pt - 1)
=\gamma_0\int_0^{w} \frac{dy}{y}(\cos y - 1).
\eea
There are three cases to be considered, depending on the range of $t$:
\begin{itemize}
\item when $t\ll1$, which corresponds to the short time behavior and implies that $y\ll1$ and $w\ll1$ (to remember that the $v(p)$ is vanishing for $p>1$); In this case we have
\be\label{assmp2}
Z(t)=\gamma_0\int_0^{w} \frac{dy}{y}(\cos y - 1)\sim \gamma_0\int_0^{w} dy(-\frac{y}{2}+O(y^3))\ll1.
\ee
So that $Z(t)$ is well defined for $y\ll1$. Furthermore, it is vanishing as $y\rightarrow 0^+$ and we have $e^{Z(t)}\vert_{t\rightarrow 0^+}\rightarrow1$.
\item when $t\in(0,1]$; In this case we can repeat the analysis as above and easily prove that $Z(t)$ is a bounded function.
\item
when $t\in [1,\infty]$; let $p_0>0$ be the minimal value of $p$ and $u=2\o_0 p_0t$, we have
\bea
Z(t)&=&\gamma_0\int_u^{2\omega_0 t} \frac{dy}{y}(\cos y - 1)\nn\\
&=&-C-\ln u-\int_0^u\frac{\cos y-1}{y} dy-[\ln 2\omega_0 t-\ln u  ]\nn\\
&=&\gamma_0(-\ln 2\omega_0 t-C-\int_0^u\frac{\cos y-1}{y} dy),\label{canc}
\eea
where we have used the integral formula \be
\int_u^{\infty}\frac{\cos y}{y}dy=-C-\ln u-\int_0^u\frac{\cos y-1}{y} dy,
\ee
where $C=0.577215\cdots$ is the Euler constant and $\int_0^u\frac{\cos y-1}{y} dy$ is a bounded function. Remark that \eqref{canc} is well defined for $u\rightarrow 0$, due to the cancellation of $\ln u$.

So we have
\be
e^{Z(t)}\sim cst\cdot [\frac{1}{2\omega_0 t}]^{\gamma_0},\quad for\  t\ge1.
\ee
\end{itemize}

Now we derive the asymptotic formula for $Q_a$ and $Q_b$. Replacing the discrete sum over $p$ in \eqref{qab} by
integrals and performing the integrations, we can easily find that: \be Q_a= Q_b=\ln\frac{(x-z)^2-t^2}{(x-z)^2-\omega_0^2t^2}.\ee

Collecting all the above terms we have: \bea \label{finalbalb} &&\lim_{L\to\io}
<I_{\l,t}|n(z)|I_{\l,t}> \nn\\
&&\quad= \frac{1}{4\pi^2}\ [{1\over( (x-z)-t)^2}+{1\over( (x-z)+t)^2}]\nn\\
&&\quad\quad\quad+
\frac{1}{2\pi^2}\ {\cos2p_F(x-z)\ e^{Z(t)}\over
(x-z)^2-(\omega_0t)^2} .\eea

So we proved theorem \ref{mainthm}.

\section{The Bosonization method}
While the Lieb-Mattis method for solving Luttinger model is mathematically rigorous, technically it is very complicated. There exist another very popular method for studying the one dimensional interacting Fermions models, called the Bosonizations, which states that certain two dimensional models of fermions are equivalent to the corresponding Bosonic models: the corresponding Fermionic Hilbert space and the Bosonic one are isomorphic and the  
the Fermionic operator can be expressed in terms of the Bosonic operators. While the Bosonization method can reduce significantly the difficulty for the calculation, it has the reputation of not mathematically rigorous. A Rigorous proof of Bosonization formulas was given very recently in a paper by Langmann and Moosavi \cite{langman}. In this section we shall prove Theorem \ref{mainthm} with the exact Bosonization formulas in \cite{langman}.
This can be considered as a verification of the use of Bosonization formula in the non-equilibrium setting.
\vskip.5cm
First of all we shall derive Formula \eqref{finalbalc}. Following the notations in \cite{langman} we have
\begin{proposition}
Let  $\rho_\o$ be the Bosonic operators introduced before and let $R^{\e_\o}_\o$ be the Klein factor, then we can express the
Fermionic operators $\psi^-$ in terms of the Bosonic operators and the Klein factor as follows: 
\bea
\psi^-_\o(x,\delta)&=&:N_\d e^{i\pi\e_\o x Q_\o/L} R^{-\e_\o}_\o e^{i\pi\e_\o x Q_\o/L}\times\\
&&\quad\quad\exp \big\{\e_\o\sum_{p> 0}  \frac{2\pi}{Lp} [\rho_\o(p)e^{-ipx-\delta |p|}-\rho_\o(-p)e^{ipx-\delta |p|} \big\}\nn,
\eea
where $\o, \o'=1, 2$, $\e_1=+$, $\e_2=-$, $Q_\o=\rho_\o(0)$ and $N_\delta=\bigg[\frac{1}{L(1-e^{-2\pi\delta/L})}\bigg]^{1/2}$ is the normalization factor. $R_\o^{\pm}$ is the Klein factor such that $R_\o^{-}=(R_\o^{+})^\dagger$. They
obey the following commutation relation (see \cite{langman} for the detailed derivation):
\bea\label{klein}
&&[\rho_\o(p), R_{\o'}]=  \e_\o\delta_{\o, \o'}\delta_{p,0} R_\o,\quad [H_0, R_\o]=\e_\o\frac{\pi}{L}\big\{\rho_\o(0), R_\o\big\},\\
&& \langle 0| R_\o^{q_1} R_{\o'}^{q_2}|0\rangle=\delta_{\o, \o'}\delta_{q_1,0}\delta_{q_2,0},\quad  R_1^{q_1} R_{2}^{q_2}=(-1)^{q_1q_2}R_{2}^{q_2}R_1^{q_1},\nn\\
&&[Q_\o, R_1^{q_1} R_{2}^{q_2}]=q_\o R_1^{q_1} R_{2}^{q_2},\quad q_\o\in\ZZZ\ . \nn
\eea
\end{proposition}
We shall not repeat the proof here and the interested reader is invited to look at \cite{langman} for details.

Let $\hat Z^-_\o=e^{i\pi\e_\o x Q_\o/L} R^{-\e_\o}_\o e^{i\pi\e_\o x Q_\o/L}$ and $\hat Z^+$ be its adjoint, we can write
the Fermionic operators as:
\be
\psi^\pm_\o(x,\delta)=N_\d\hat Z^\pm_\o e^{\mp\e_\o\sum_{p> 0}  \frac{2\pi}{Lp} [\rho_\o(p)e^{-ipx-\delta |p|}-\rho_\o(-p)e^{ipx-\delta |p|} }.
\ee

We calculate first the term $\langle0|\ \psi_{1, x}^-
e^{i H t} \psi_{1, z}^+ \psi_{2, z}^- e^{-i H t}
\psi_{2, x}^+\  |0 \rangle$ in\eqref{wickord} forget the phase
factor $e^{ip_F(x-z)}$ for the moment. Inserting the identity operators $I=e^{iHt}e^{-iHt}$
and $I=e^{i S}e^{-i S}$ we derived Formula \eqref{id1}, which is
the starting point of our analysis.

First of all, it is easily to find that \be e^{i S}\hat Z_\o^{\pm}e^{-i
S}=\hat Z_\o^{\pm}.\ee 
Using Formula \eqref{rho} we have:
\bea
e^{iS}\psi_{1,z}^+e^{-iS}&=&N_\delta \hat Z_1^{+}
\exp\frac{2\pi}{L}\sum_{p>0}\frac{1}{p} \
\big\{-e^{-\delta p}e^{-ipz} [ \cosh\phi\rho_1(p)+\sinh\phi\rho_2(p)]\nonumber\\
&&\quad\quad\quad+ e^{-\delta
p}e^{ipz}[\cosh\phi\rho_1(-p)+\sinh\phi\rho_2(-p)]\big\},\nn \eea
and \bea e^{iS}\psi_{2,z}^-e^{-iS}&=&N_\delta\hat Z_2^-
\exp\frac{2\pi}{L}\sum_{p>0}\frac{1}{p} \
\big\{-e^{-\delta p}e^{-ipz} [ \cosh\phi\rho_2(p)+\sinh\phi\rho_1(p)]\nonumber\\
&&\quad\quad\quad+ e^{-\delta
p}e^{ipz}[\cosh\phi\rho_2(-p)+\sinh\phi\rho_1(-p)]\big\}. \eea

Using the fact that:
\be
[ H_0+D,\ R^{\pm}_\o]=\pm \frac{2\pi(\s(0)+1)}{L}\ R^{\pm}_\o(2\e_\o\rho_\o (0)+1),
\ee
and
\bea
&&e^{i(H_0+D)t}R^{\pm}_\o e^{-i(H_0+D)t}\nn\\
&&\quad\quad=R^{\pm}_\o \exp\ [\pm
\frac{2\pi(\s(0)+1)}{L}(2\e_\o\rho_\o (0)+1)t\ ], \eea 
we have:
%
\bea
&& e^{-i S} e^{i (H_0+D) t}e^{iS}\psi_{1,z}^+e^{-iS}  e^{i (H_0+D) t}e^{i S}\\
&=&N_\delta\hat Z_1(t) \exp\frac{2\pi}{L}\sum_{p>0}\frac{1}{p} \
\big\{e^{-\delta p} [ A_{1}\rho_1(p)+ A_{-1}\rho_1(-p)+ A_{2}\rho_2(p)+A_{-2}\rho_2(-p)]\big\},
\nonumber\eea
and
\bea
&& e^{-i S} e^{i (H_0+D) t}e^{iS}\psi_{2,z}^-e^{-iS}  e^{i (H_0+D) t}e^{i S}\\
&=&N_\delta\hat Z_2(t)  \exp\frac{2\pi}{L}\sum_{p>0}\frac{1}{p} \
\big\{e^{-\delta p} [ B_{1}\rho_1(p)+ B_{-1}\rho_1(-p)+ B_{2}\rho_2(p)+B_{-2}\rho_2(-p)]\big\},\nonumber
\eea
where
\bea
&&A_{\pm1}=\pm e^{\mp ip[z+(\s+1)t]}\ \sinh^2\phi\mp e^{-ip[z-(\s+1)t]}\ \cosh^2\phi\ ,\nonumber\\
&&A_{\pm2}=\pm e^{\mp ip[z- (\s+1)t]}\ \sinh\phi\cosh\phi\mp e^{\mp ip[z+ (\s+1)t]}\ \cosh\phi\sinh\phi ,\nn\\
&&B_{\pm1}=\pm e^{\mp ip[z+ (\s+1)t]}\sinh\phi\cosh\phi\mp e^{\mp ip[z- (\s+1)t]}\cosh\phi\sinh\phi ,\nonumber\\
&&B_{\pm2}=\pm e^{\mp ip[z- (\s+1)t]}\sinh^2\phi\mp e^{\mp ip[z+ (\s+1)t]}\cosh^2\phi \nn\\
&&\hat Z_1(t)=e^{i\pi x \rho_1(0)/L} \exp[-\frac{2\pi(\s(0)+1)}{L}(2\rho_1 (0)+1)\ t]R^{-1}_1 e^{i\pi x \rho_1(0)/L},\nn\\
&&\hat Z_2(t)=e^{i\pi x \rho_2(0)/L} \exp\{\frac{2\pi(\s(0)+1)}{L}(-2\rho_2
(0)+1)t\}R_2 e^{i\pi x \rho_2(0)/L}.
\eea

When $p=0$, by using the fact that
$\rho_\o(0) |0\rangle=0$ and $\langle0|R_\o^{q_1} R_{\o'}^{q_2}|0\rangle=\delta_{\o, \o'}\delta_{q_1,0}\delta_{q_2,0}$,
we have
\be
\langle 0|\hat Z_1 \hat Z^+_1(t) \hat Z_2(t)\hat Z^\dagger_2|0\rangle=1.
\ee
So the nontrivial contributions come from the $p>0$ part. Using repeatedly the Hausdorff formula
we can factorize the terms depending on $\r_1(\pm p)$ and $\r_2(\pm p)$:
\bea
&&\langle0| \ N_\delta \exp {\frac{2\pi}{L}\sum_{p>0}\frac{1}{p}\ [ e^{-\delta p}e^{-ipx}\rho_1(p)- e^{-\delta p}e^{ipx}\rho_1(-p)]}\nonumber\\
&&\times N_\delta \exp\frac{2\pi}{L}\sum_{p>0}\frac{1}{p} \
\big\{e^{-\delta p} [ A_{+1}\rho_1(p)+ A_{-1}\rho_1(-p)+A_{+2}\rho_2(p)+A_{-2}\rho_2(-p)]\big\}\nonumber\\
&&\times N_\delta \exp\frac{2\pi}{L}\sum_{p>0}\frac{1}{p} \
\big\{e^{-\delta p} [ B_{+1}\rho_1(p)+ B_{-1}\rho_1(-p)+ B_{+2}\rho_2(p)+B_{-2}\rho_2(-p)]\big\}\nonumber\\
&&\times N_\delta e^{\frac{2\pi}{L}\sum_{p>0}\frac{1}{p} \
[e^{-\delta p}e^{ipx}\rho_2(p)- e^{-\delta p}e^{-ipx}\rho_2(-p)]}\ |0\rangle\nonumber\\
&&=:N_\delta^4\ I_1\ I_2, \eea where \bea I_1&=&\langle
0|e^{\frac{2\pi}{L}\sum_{p>0}\frac{1}{p}\ \  e^{-\delta
p}[e^{-ipx}\rho_1(p)- e^{ipx}\rho_1(-p)]}\
e^{\frac{2\pi}{L}\sum_{p>0}\frac{1}{p} \
e^{-\delta p} [ A_{+1}\rho_1(p)+ A_{-1}\rho_1(-p)]}\nonumber\\
&\times& e^{\frac{2\pi}{L}\sum_{p>0}\frac{1}{p} e^{-\delta p} [ B_{+1}\rho_1(p)+ B_{-1}\rho_1(-p)]}\ |0\rangle,\\
I_2&=&\langle 0| e^{\frac{2\pi}{L}\sum_{p>0}\frac{1}{p} \
e^{-\delta p} [ A_{+2}\rho_2(p)+ A_{-2}\rho_2(-p)]}\ \  e^{\frac{2\pi}{L}\sum_{p>0}\frac{1}{p} e^{-\delta p} [ B_{+2}\rho_2(p)+ B_{-2}\rho_2(-p)]}\nonumber\\
&\times& e^{\frac{2\pi}{L}\sum_{p>0}\frac{1}{p} \
e^{-\delta p}[e^{-ipx}\rho_2(p)-e^{ipx}\rho_2(-p)]} |0\rangle.
\eea
Following exactly the same procedure as section \ref{calci}, namely using repeatedly the Hausdorff formula and the annihilation formulas we
have:
\bea\label{t1}
I_1\ I_2&=&\exp\frac{2\pi}{L}\sum_{p>0}\frac{1}{p} \  e^{-2\delta p}[(e^{ip(x-z)+ ip (\s+1)t}-1)+(e^{ip(x-z)- ip (\s+1)t}-1)\nn\\
&+& 2\sinh\phi\cosh\phi(\sinh\phi+\cosh\phi)^2(\cos 2p(\s+1)t-1)].
\eea

In order to reproduce the expressions in \eqref{finalbalc} we need to extract from the above formula the noninteracting 2-point correlation function (see \cite{langman}), as follows. We write the terms
$e^{\pm ip(x-z)\pm ip (\s+1)t}-1$ in the above formula as $$(e^{ ip(x-z)\pm ip (\s+1)t}-e^{ ip(x-z)\pm ip (\s+1)t})+(e^{ ip(x-z)\pm ip (\s+1)t}-1),$$
while the first term gives the factors $Q$, the second term contributes to the non-interacting correlation function:
\bea N_\delta^4\exp\frac{2\pi}{L}\sum_{p>0}\frac{1}{p} \
e^{-2\delta p} [(e^{ip(x-z)+ ip t}-1)+(e^{-ip(x-z)+ ip t}-1)].\label{nonintb}
\eea 

Now we derive the asymptotic formula for \eqref{nonintb}. Using the Poisson summation
formula: \bea \exp \bigg( \sum_{p>0}\frac{2\pi}{Lp}e^{-2\delta p}
\bigg)=\exp\bigg( \sum_{n=1}^{\infty}\frac{1}{n}e^{-4n\pi\delta
/L}  \bigg)=LN_\delta^2, \eea where \be N_\delta^2=
\frac{1}{L(1-e^{-4\pi\delta/L})}\sim \frac{1}{4\pi\delta}  \ for \
{L\rightarrow\infty}, \ee
Formula \eqref{nonintb} can be written as:
 \bea &&\lim_{\d\rightarrow
0^+}\lim_{L\rightarrow\infty}N_\delta^4\exp\frac{2\pi}{L}\sum_{p>0}\frac{1}{p}
\
e^{-2\delta p} [(e^{ip(x-z)+ ip t}-1)+(e^{-ip(x-z)+ ip t}-1)]\nn\\
&=&\lim_{\d\rightarrow 0^+}\lim_{L\rightarrow\infty}N_\delta^4\
\frac{1}{L^2N_\delta^4}\cdot\frac{1}{1-e^{-\frac{2\pi}{L}[2\delta
+i(x-z)+ i t]}}\ \frac{1}{1-e^{-\frac{2\pi}{L}[2\delta -i(x-z)+ i
t]}}\nn\\
&=&\frac{1}{4\pi^2}\frac{1}{(x-z)^2- t^2}. \eea



Following the same procedure we can calculate all the other terms in \eqref{wickord} and
derive Formula \eqref{finald}. The asymptotic expressions for the terms in the exponential can be derived with the same procedure as in the last section and we shall not repeat it here. So we proved
Theorem \eqref{mainthm} with the exact Bosonization formulas.

{\bf Acknowledgement} 
The author is very grateful to V. Mastropietro for many useful discussions and common work on the preliminary version of this paper. Part of this work has been done during the author's visit to the Chern Institute of Technology, Nankai University and the Shanghai Institute of Mathematical Science, Fudan University. The author is very grateful for their hospitality. 
The work is partially supported by the National Natural Science Foundation of China under the grant No. 11701121.
\section{Appendix }
\subsection{Explicit expressions of the factors in Formulas \eqref{i1} and \eqref{i2}}

With some very long but elementary calculation we find that
the expressions of the terms in formula \eqref{i1} and \eqref{i2} read:
\bea\label{za}
&&z_a=\exp\frac{2\pi}{L}\sum_p\frac{1}{p}\{(e^{-ip\s t}-1)\},\\
&&A_{1\pm}=\exp \frac{2\pi}{L}\sum_p\frac{1}{p}\rho_1(\pm p)\cosh\e\phi (\mp e^{\mp ipx\pm ip t}\pm e^{\mp ipx\pm ip t(\s+1)})\nonumber\\
&&A_{2\pm}=\exp\pm \frac{2\pi}{L}\sum_p\frac{1}{p}\rho_2(\pm p)\sinh\e\phi (e^{\mp ipx\pm ip t}-e^{\mp ipx\pm ip t(\s+1)}).\nn
\eea
\bea\label{2WR2}
&&z_b=\exp \frac{2\pi}{L}\sum_p\frac{1}{p}(1-e^{-ip\sigma t})=z_a^{-1},\\
&&B_{1\pm}=\exp \mp \frac{2\pi}{L}\sum_p\frac{1}{p}\rho_1(\pm p)\sinh\e\phi (\mp e^{\mp ipz\mp ipt}\pm e^{\mp ipz\mp ipt(\s+1)})\ ,\nonumber\\
&&B_{2\pm}=\exp \frac{2\pi}{L}\sum_p\frac{1}{p}\rho_2(\pm p)\cosh\e\phi (\pm e^{\mp ipz\mp ipt(\s+1)}\mp e^{\mp ipz\mp ipt})\ .\nonumber
\eea
\bea\label{1pt1}
\tilde W^{-1}_{1t\e}&=&\exp \frac{2\pi}{L}\sum_p\frac{1}{p} [ (\cosh\e\phi-1) e^{-ipz+ipt}\rho_1(p)- (\cosh\e\phi-1) e^{ipz-ipt}\rho_1(-p)\  ]\ ,\nonumber\\
\tilde R^{-1}_{1t\e}&=&\exp -\frac{2\pi}{L}\sum_p\frac{1}{p} [\sinh\e\phi e^{-ipz+ipt}\rho_2(p)-\sinh\e\phi e^{ipz-ipt}\rho_2(-p)\  ].
\eea
\bea
\bar W^{-1}_{1t\e}&=&\exp \frac{2\pi}{L}\sum_p\frac{1}{p} [ (\cosh\phi-1)\cosh\e\phi e^{-ipz+ip(\s+1)t}\rho_1(p)\\
&-& (\cosh\phi-1)\cosh\e\phi e^{ipz-ip(\s+1)t}\rho_1(-p)\  ]\ ,\nonumber\\
\bar R^{-1}_{1t\e}&=&\exp -\frac{2\pi}{L}\sum_p\frac{1}{p} [ (\cosh\phi-1)\sinh\e\phi e^{-ipz+ip(\s+1)t}\rho_2(p)\nonumber\\
&-& (\cosh\phi-1)\sinh\e\phi e^{ipz-ip(\s+1)t}\rho_2(-p)\  ]\nonumber\\
\hat W^{-1}_{1t\e}&=&\exp -\frac{2\pi}{L}\sum_p\frac{\sinh\phi}{p}[\rho_1(p)\sinh\e\phi e^{-ipx-ipt(\s+1)}\nonumber\\
&-&\rho_1(-p)\sinh\e\phi e^{ipx+ipt(\s+1)}],\nonumber\\
\hat R^{-1}_{1t\e}&=& \exp -\frac{2\pi}{L}\sum_p\frac{\sinh\phi}{p}[\ \rho_2(-p)\cosh\e\phi e^{ipx+ipt(\s+1)}\nonumber\\
&-& \rho_2(p)\cosh\e\phi e^{-ipx-ipt(\s+1)}]\nonumber\ .
\eea

\bea
&&\tilde W_{2t}=\exp -\frac{2\pi}{L}\sum_p\frac{1}{p} \ \sinh\e\phi (e^{-ipz-ipt}\rho_1(p) -e^{ipz+ipt}\rho_1(-p)),\\
&&\tilde R_{2t}=\exp\frac{2\pi}{L}\sum_p\frac{1}{p}\   (\cosh\e\phi-1) ( e^{-ipz-ipt}\rho_2(p)- e^{ipz+ipt}\rho_2(-p)). \nonumber
\eea
\bea
\bar W_{2t\e}&=&\exp \frac{2\pi}{L}\sum_p\frac{1}{p}  \sinh\phi\cosh\e\phi\ [ e^{-ipz+ip(\s+1)t}\rho_1(p)\\
&-& e^{ipz-ip(\s+1)t}\rho_1(-p)\  ],\nn\\
\hat W_{2t\e}&=&\exp -\frac{2\pi}{L}\sum_p\frac{1}{p}  (\cosh\phi-1)\sinh\e\phi [\rho_1(p) e^{-ipz-ipt(\s+1)}\nn\\
&-&\rho_1(-p) e^{ipz+ipt(\s+1)}],\nonumber\\
\bar R_{2t\e}&=& \exp -\frac{2\pi}{L}\sum_p\frac{\sinh\phi\sinh\e\phi}{p}[\ \rho_2(p) e^{-ipz+ipt(\s+1)}\nn\\
&-&\rho_2(-p) e^{ipz-ipt(\s+1)}],\nonumber\\
\hat R_{2t\e}&=& \exp \frac{2\pi}{L}\sum_p\frac{1}{p}(\cosh\phi-1)\cosh\e\phi\ [\ \rho_2(p)  e^{-ipz-ipt(\s+1)}\nn\\
&-& \rho_2(-p) e^{ipz+ipt(\s+1)}\ ].\nonumber
\eea

\end{document}